\documentclass[useAMS,usenatbib,usegraphicx]{mn2e}

\def\ltsima{$\; \buildrel < \over \sim \;$}
\def\simlt{\lower.5ex\hbox{\ltsima}}
\def\gtsima{$\; \buildrel > \over \sim \;$}
\def\simgt{\lower.5ex\hbox{\gtsima}}
\def\gsimeq
{\hbox{\raise0.5ex\hbox{$>\lower1.06ex\hbox{$\kern-1.07em{\sim}$}$}}}
\def\lsimeq
{\hbox{\raise0.5ex\hbox{$<\lower1.06ex\hbox{$\kern-1.07em{\sim}$}$}}}

\def\xmm{{\it XMM-Newton }}

\def\xmm{{\it XMM-Newton}}
\def\chandra{{\it Chandra}}
\def\suzaku{{\it Suzaku}}
\def\rxte{{\it RXTE}}

\def\apj{ApJ}
\def\mnras{MNRAS}
\def\aap{A\&A}
\def\apjl{ApJ}
\def\apjs{ApJS}
\def\araa{ARA\&A}

\def\nat{Nature}

\def\grsdn{GRS1915+105}

\def\gros{GROJ1655-40}
\def\hs{H1743-322}
\def\gxt{GX339-4}
\def\xtejd{XTEJ1817-330}
\def\us{4U1630-47}
\def\ud{4U1957+115}

\def\Fevc{Fe {\sc xxv}}
\def\Fevs{Fe {\sc xxvi}}

\def\tic{T$_{\rm IC}$}

\def\xis{XIS}
\def\xis1{XIS1}
\def\xis2{XIS2}
\def\xis3{XIS3}

\title[] 
 {{Ubiquitous equatorial accretion disc winds in black hole soft states}}

 \author[G.\ Ponti et al. ]
 {G.~Ponti$^{1}$,
R. P. Fender$^{1}$, M. C. Begelman$^{2,3}$, R. J. H. Dunn, J. Neilsen$^{4}$ and M. Coriat$^{1}$ 
\\ \\
   $^1$School of Physics and Astronomy, University of Southampton, Highfield, Southampton, SO17 1BJ, UK\\
   $^2$JILA, University of Colorado and National Institute of Standards and Technology, Boulder, CO 80309-0440, USA \\
   $^3$Department of Astrophysical and Planetary Sciences, University of Colorado, Boulder, CO 80309-0391, USA \\
   $^4$MIT Kavli Institute for Astrophysics and Space Research, Cambridge, MA 02139\\
}

\pagerange{\pageref{firstpage}--\pageref{lastpage}}

\usepackage{times}

\begin{document}

\label{firstpage}

 \maketitle

\begin{abstract}
High resolution spectra of Galactic Black Holes (GBH) reveal the presence 
of highly ionised absorbers. In one GBH, accreting 
close to the Eddington limit for more than a decade, a powerful 
accretion disc wind is observed to be present in softer X-ray states 
and it has been suggested that it can carry away enough mass and 
energy to quench the radio jet.
Here we report that these winds, which may have mass outflow rates 
of the order of the inner accretion rate or higher, are an ubiquitous 
component of the jet-free soft states of all GBH. We furthermore demonstrate 
that these winds have an equatorial geometry with opening angles of few 
tens of degrees, and so are only observed in sources in which the 
disc is inclined at a large angle to the line of sight. 
The decrease in \Fevc/ / \Fevs\ line ratio with Compton temperature, 
observed in the soft state, suggests a link between higher wind ionisation
and harder spectral shapes.  Although the physical interaction 
between the wind, accretion flow and jet is still not fully understood, 
the mass flux and power of these winds, and their presence ubiquitously 
during the soft X-ray states suggests they are fundamental components of the 
accretion phenomenon. 
\end{abstract}


\section{Introduction}

The feedback of liberated gravitational potential energy by accreting
black holes is determined by the combination of accretion states and
outflow modes. In galactic black holes (GBH), hysteresis is observed
between the X-ray state of the accretion flow, which is strongly
coupled to the presence of a relativistic jet, and the luminosity of
the source (Fender et al. 2004). The jet is always present in 
`hard' X-ray states, which can be observed at all luminosities. 
However, at the highest luminosities sources can enter into a `soft' 
X-ray state in which the jet is switched off and kinetic feedback therefore 
appears to be strongly suppressed. Once the soft state is entered, 
GBH remain in this state until they decline to $\sim 1$\% of the Eddington 
rate, at which point they return to the hard state. Interestingly, it has been 
suggested that similar states also apply to AGN (Koerding et al. 2006).

Recent high energy resolution observations of several GBH showed 
the presence of winds (Lee et al. 2002; Miller et al. 2004; 2006a,b), 
indicating that these objects drive outflows not only in the form of jets, 
but also of winds (Diaz-Trigo et al. 2011).
There are three main mechanisms that can launch a wind from the
surface (the atmosphere) of the accretion disc: thermal, radiation and
magnetic pressure. In each case, a wind will be launched only if the
pressure can overcome gravity. As a rule of thumb, the closer the 
launching point is to the BH, the higher the wind terminal velocity.

In GRS1915+105, a peculiar GBH accreting close to the Eddington rate 
for more than a decade (Fender \& Belloni 2004), an accretion disc 
wind appears to be present in softer X-ray states and to be so powerful 
and to carry away so much mass as to halt the flow of matter into 
the jet (Neilsen \& Lee 2009). 

\section{Wind-state-jet connection}

\begin{figure*}
\includegraphics[width=0.75\textwidth,height=1.0\textwidth,angle=-90]{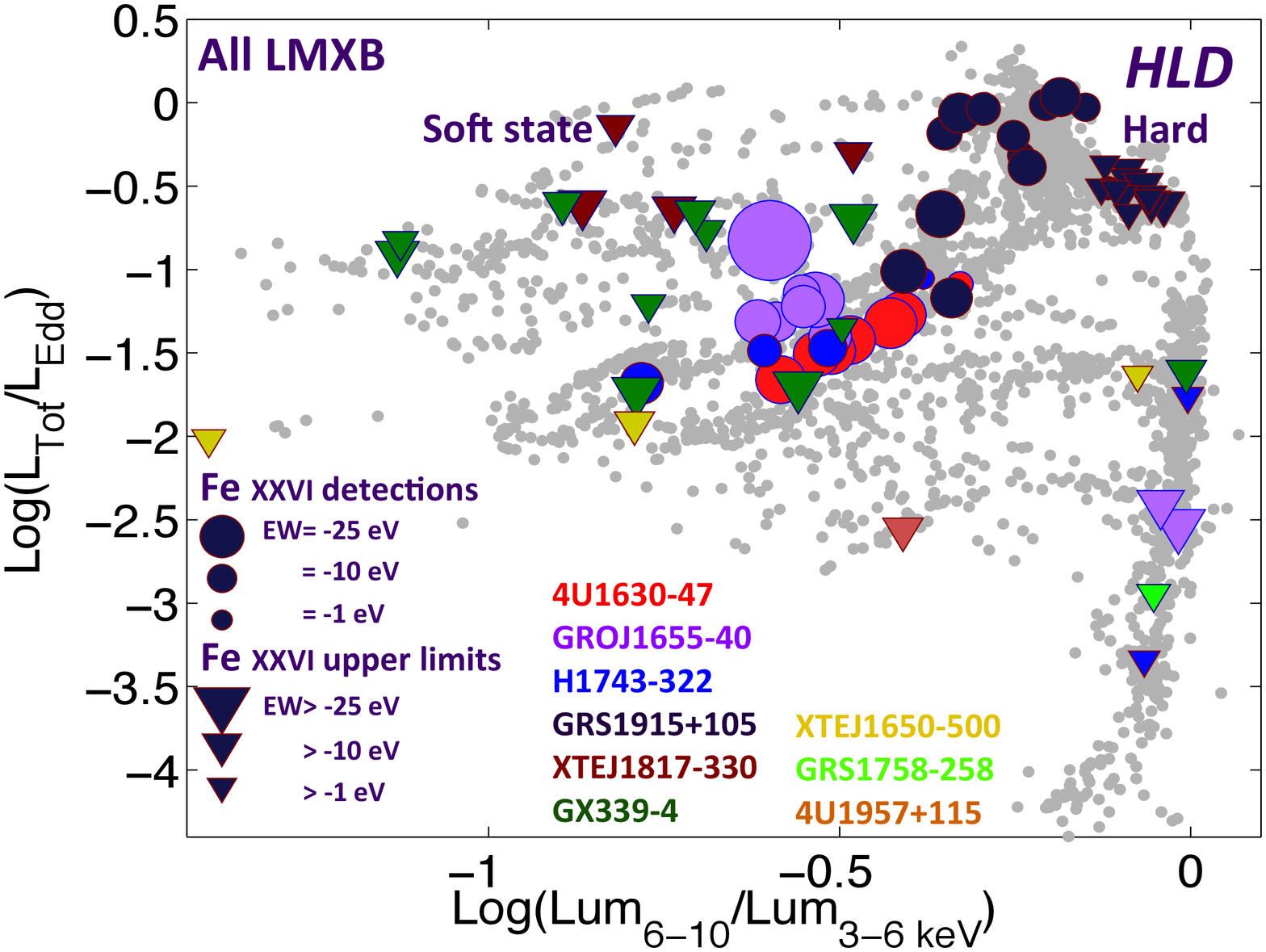}
\caption{The presence or absence of ionised winds in a sample of GBH.
The grey (background) points show the Hardness Luminosity 
Diagram (HLD) of all Low Mass X-ray Binaries (LMXB) studied. 
The hardness is computed from \rxte\ data as Log($\frac{L_{6-10}}{L_{\rm 3-6}}$), 
where L$_{\rm 6-10}$ and L$_{\rm 3-6}$ are the source luminosity 
in the 6-10 and 3-6 keV bands, respectively. 
Circles indicate \chandra, \xmm\ and \suzaku\ observations during which 
a high ionisation wind is detected through the observation of the \Fevs\ 
absorption line. The size of the symbol is directly proportional to the EW 
of the \Fevs\ absorption line. Triangles show, instead, the non-detections. 
Their size is proportional to the upper limit on the \Fevs\ EW.
The different colors indicate data results from different sources. 
In soft states, during which the jet is always quenched, sometimes 
a wind is observed. However, other soft state observations show 
stringent upper limits on the presence of the wind. 
}
\label{DFLDall}
\end{figure*}

We have performed a comprehensive study of ionised X-ray winds in GBH.
To do this, we analysed the X-ray spectra of all the \chandra, \xmm\ and \suzaku\ 
observations of black hole Low Mass X-ray Binaries (LMXB, which are GBH 
accreting by Roche lobe overflow) with well studied outbursts (Dunn et al. 2010) 
and at least one deep (exposure $>5$~ks) grating spectroscopy observation. 
An effective way to separate the observations during the soft and hard states 
is to plot the Hardness Luminosity Diagram (HLD). 
We compute the HLD from data obtained with the Rossi X-ray Timing Explorer 
(\rxte) by first fitting each spectrum of each source with two 
components, a multi-temperature disc black body component for the disc 
and a power law component for the corona (see Dunn et al. 2010 for more details). 
The grey (background) points in Fig. \ref{DFLDall} show the total luminosity (in 
Eddington units) vs. the hardness for each RXTE observation. 
The hardness is computed as Log($\frac{L_{6-10}}{L_{\rm 3-6}}$),
where L$_{\rm 6-10}$ and L$_{\rm 3-6}$ are the observed source luminosity 
in the 6-10 and 3-6 keV bands, respectively. During observations in 
the hard state, the spectrally hard power law component dominates 
the emission, thus Log($\frac{L_{6-10}}{L_{\rm 3-6~keV}}$) is close to 0. 

Circles in Fig. \ref{DFLDall} correspond to each \chandra, \xmm\ and 
\suzaku\ observations during which a high ionisation wind is detected 
through the observation of the \Fevs\ absorption line. In particular 
the size of the symbol is directly proportional to the EW of the \Fevs\ 
absorption line. Triangles, instead, report wind non-detections 
(symbol size proportional to the \Fevs\ upper limit). 

Analysing just 11 GRS1915+105 \chandra\ observations, 
Neilsen \& Lee (2009) already measured a strong anti-correlation between 
the presence of the (radio) jet and of the winds, with the wind being present 
primarily during the jet--free soft states and disappearing during hard states
(see also Miller et al. 2008). 
The addition of the \xmm\ and \suzaku\ data, which more than doubles the 
number of observations, confirms and strengthens the anti-correlation (with 
now a total of 26 good quality spectra) in this peculiar LMXB. 

However, when we consider other LMXBs, the picture appears to be 
more complex. In fact, stringent upper limits are set during 17 soft 
state (jet-quenched) observations. In the next section, we suggest 
that this may be an effect of viewing geometry.

\section{The wind angular dependence}

\begin{figure*}
\includegraphics[width=0.46\textwidth,height=0.49\textwidth,angle=-90]{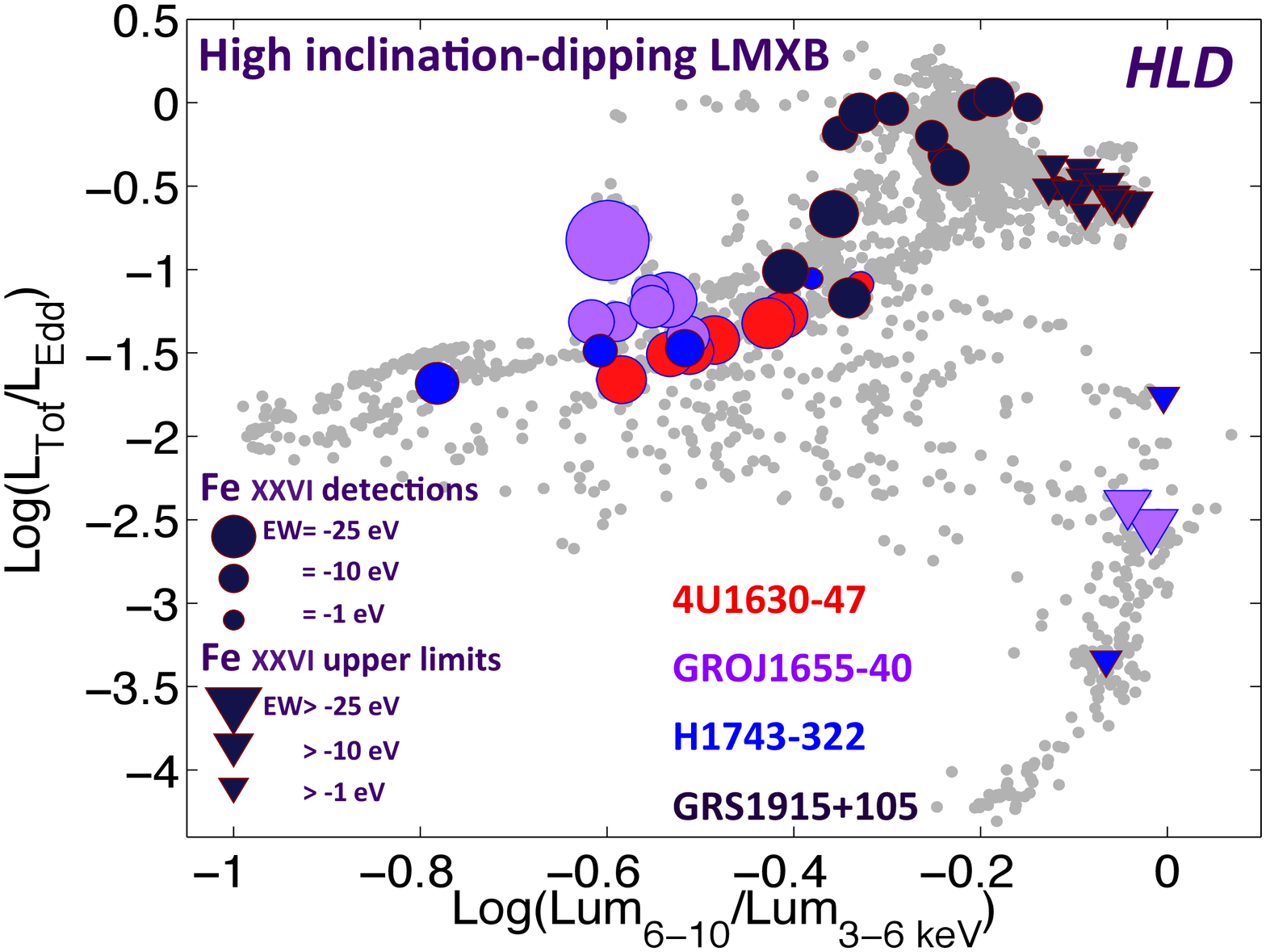}
\includegraphics[width=0.46\textwidth,height=0.49\textwidth,angle=-90]{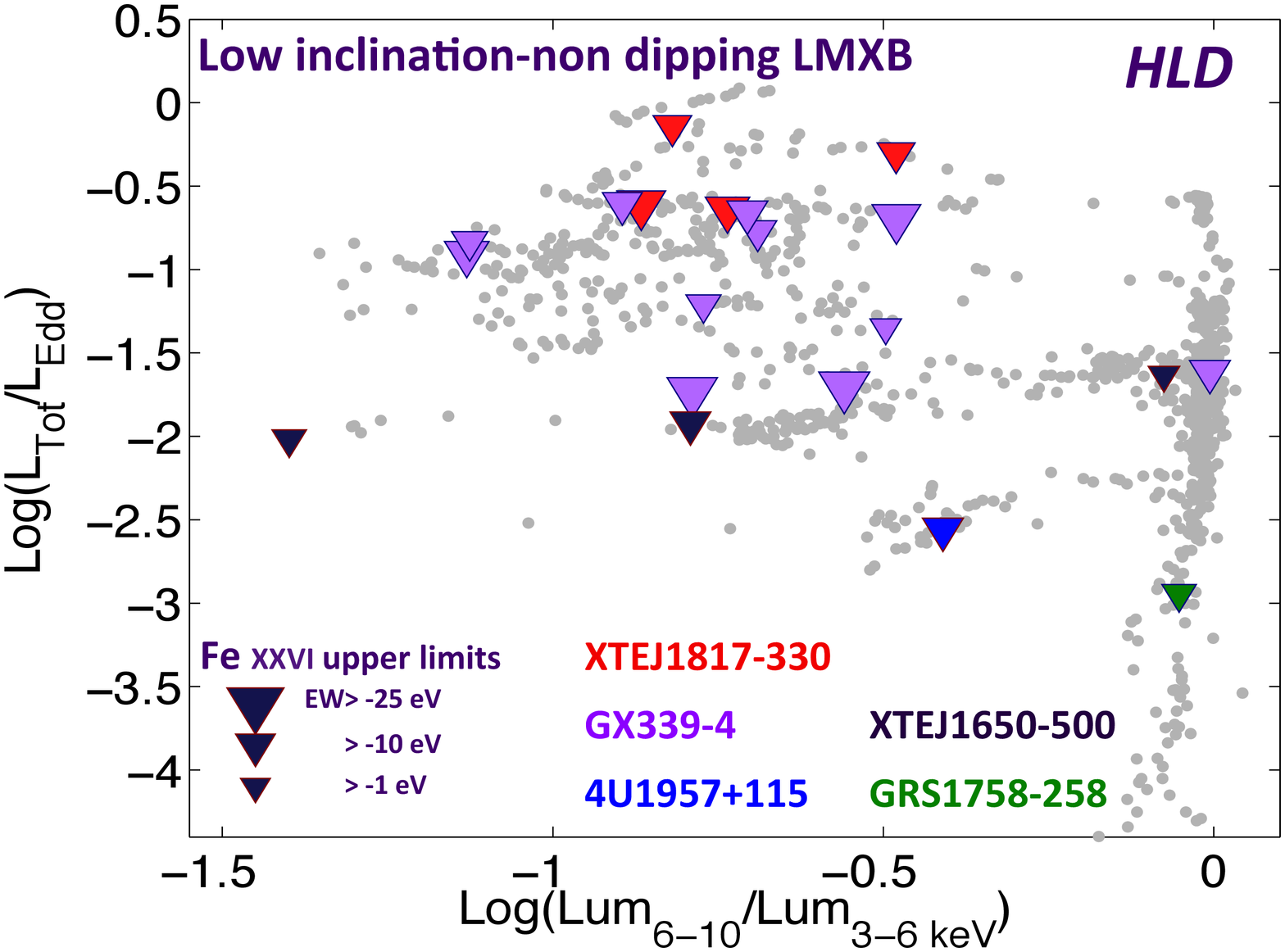}
\caption{{\it (Left panel)} HLD of the high inclination (dipping) LMXB studied and 
of all the low inclination (non dipping) LMXB, {\it right panel}. High inclination (dipping) sources show \Fevs\ absorption every time they are in the soft state and upper limits in the hard states. In low inclination (non dipping) LMXB the \Fevs\ absorption line is never detected. We interpret these as due to an ubiquitous equatorial disc wind associated with soft states only. We note that high inclination sources tend to show a more triangular HLD, while the low inclination sources exhibit a boxy one.}
\label{DFLDDip}
\label{DFLDNoDip}
\end{figure*}

To study the wind angular dependence we aim at dividing the sources 
into two samples, based on inclination.
Thanks to extensive campaigns at other wavelengths, we know that 
\gros\ and \grsdn\ are high inclination sources, close to edge-on, with 
similar disc inclinations of $\sim70$ degrees (van der Hooft 1998; 
Greiner et al. 2001). 
\gros\ is also known to experience frequent dips (Tanaka et al. 2003). 
The dipping phenomenon is thought to be produced by clumps of low 
ionisation material along the line of sight, that are temporarily obscuring 
the X-ray source. The intervening material is probably related to the 
transfer of matter from the companion star to the disc and it generally 
occurs in sources observed at high inclination (Frank, King \& Raine 2002). 
Therefore we also added \hs\ and \us, known to experience frequent 
dips (Homan et al. 2005; Tomsick et al. 1998), to represent a population 
of sources which are close to edge on.
Fig. \ref{DFLDDip} (left panel) shows the HLD of all the high
inclination LMXB and reports the measured \Fevs\ absorption line
Equivalent Width (EW). These sources show clear evidence for a high
ionisation disc wind (v$_{\rm out}\sim10^{2.5-3.5}$ km s$^{-1}$) during all
30 observations in the soft state\footnote{One observation
  of GRS1915+105 with lower luminosity and Compton temperature 
  does not show any \Fevs\ but only \Fevc\ absorption, thus suggesting 
  the importance of ionisation effects.}.  

On the other hand whenever these sources are observed in the hard
X-ray state, they show only upper limits. We, in fact, observe
stringent upper limits for 16 out of 17 observations and just one
detection of a weak wind quasi-contemporaneous with a weak jet 
(Lee et al. 2002; Neilsen \& Lee 2009). 
This demonstrate that for this set of sources the presence 
of the disc wind is deeply linked to the source state. In particular, the wind is
present during spectrally-soft states, when the jet emission is strongly
quenched.

The right panel of Fig. \ref{DFLDNoDip} shows the HLD for the non
dipping LMXB, \gxt, \xtejd, \ud, XTEJ1650-500 and GRS1758-258, which
have accretion discs which are inclined more face-on to the
observer. None of these source has a detection of a highly ionised
wind in any state. Several spectra have a signal-to-noise ratio 
good enough to measure upper limits as small as a few eV, 
even during the soft state. For this reason, we confidently state that these
sources do not present the signatures of highly ionised FeK
winds\footnote{The majority of the low energy absorption lines
  detected in these LMXB (Miller et al. 2004) are consistent with being produced by
  the interstellar medium (Nowak et al. 2004; Juett et al. 2004; 2006). 
  Most of the remaining structures are consistent with being at rest, 
  thus unlikely associated to the FeK wind (Juett et al. 2006).}.

This difference in behaviour can be easily understood if both the high and 
low inclination sources have the same wind present in soft states and absent
in the hard states, but the wind is concentrated in the plane of the disc; thus 
our line of sight intercepts the wind only in high inclination sources. 
If this idea is correct, we expect that deeper observations of low inclination 
sources may reveal the presence of the wind through the detection of weak 
ionised emission lines.

Is it theoretically plausible for the disc winds to have a strong angular 
dependence? Indirect evidence for an angular
dependence of the wind in GBH was already inferred from the lack of
emission lines associated with the X-ray absorption lines (Lee et al. 2002; 
Miller et al. 2006). This suggests that the wind subtends a small fraction 
of $4\pi$ sr.   
Moreover, disc wind models and magneto hydrodynamic simulations 
predict a strong angular dependence of the wind (Begelman et al. 1983a,b; 
Melia et al. 1991; 1992; Woods et al. 1996; Luketic et al. 2010; 
Proga et al. 2002).  
In fact, if the disc wind is produced by X-ray irradiation (i.e. Compton heating, 
line driving), it is expected to be stronger in edge-on sources simply because 
once the material is lifted from the disc, it will experience an asymmetric push 
from the radiation field of the central source.
Flattened disc winds have also been assumed to explain the winds of broad 
absorption-line QSO and other AGN outflows (e.g. Emmering et al. 1992; 
Murray et al. 1995; Elvis 2000).

\section{Ionisation effects}

The strong connection between winds and source states
requires an explanation. Ueda et al. (2010), during oscillating X-ray 
states of \grsdn, observe the ionisation parameter of the wind to 
vary with the source luminosity, suggesting the importance of ionisation. 
Can an over-ionisation effect explain the disappearance of the wind
during hard states?
If the absorber is in the form of ''static'' clouds with approximately 
constant density $n$ and distance $R$ from the ionising source 
and assuming that the spectral shape changes have a minor impact 
on the ionisation state of the wind, then the 
absorber ionisation parameter $\xi$ will be directly related to the source 
luminosity\footnote{Where $L$ has been computed as the integral of 
the disc emission (in the 0.001-100 keV band) plus the power law (1-100 keV) 
one.}: $\xi=L/n R^2$. 
The left panel of Fig. \ref{DFLDNoDip} shows that {\it at the 
same luminosity} the winds are present in the soft but not in hard 
states (see also Lee et al. 2002; Miller et al. 2006b; Blum et al. 
2010; Neilsen et al. 2011). 
Thus, the "static absorber" interpretation may be unlikely.

\begin{figure}
\begin{center}
\includegraphics[width=0.38\textwidth,height=0.52\textwidth,angle=-90]{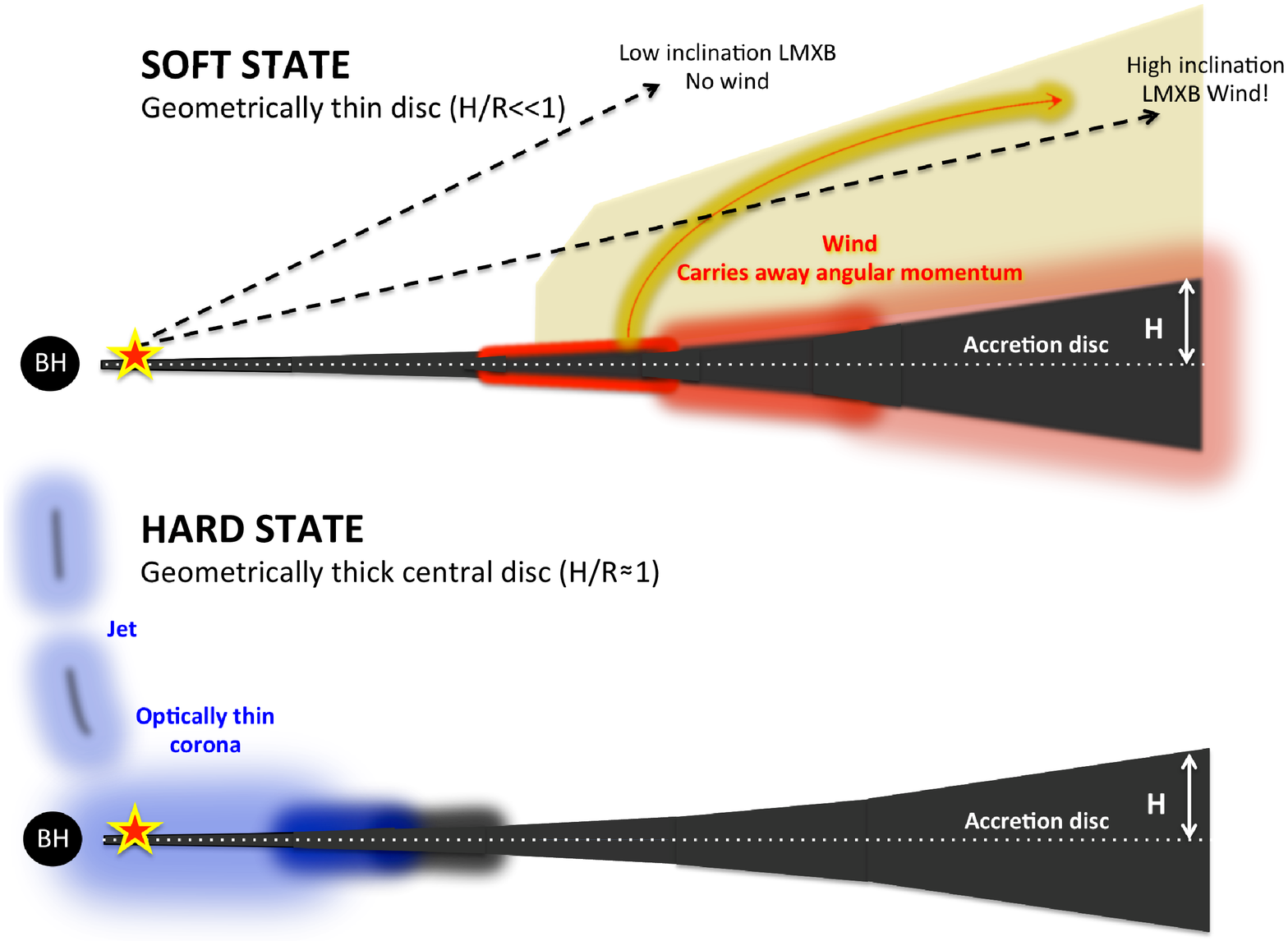}
\caption{Several physical mechanisms can explain the properties of the observed equatorial disc winds. Here the thermal winds scenario is sketched. In soft states, associated with geometrically thin discs, the central source does probably illuminate the outer disc and thus it might heat it, increasing the thermal pressure that then drives away a wind, which is flattened above the disc. Thus, only in high inclination sources our line of sight to the central source crosses the wind, allowing us to detect it. In hard states a geometrically thick and optically thin corona and the jet are present, while no wind is observed.}
\label{WindPicture}
\end{center}
\end{figure}
Early works on accretion disc theory (Shakura \& Sunyaev 1973) 
already predicted the formation of winds from the outer disc. 
Compton heated winds (see Fig. \ref{WindPicture}) can be launched 
if the inner disc is geometrically thin and thus the central source 
can illuminate and heat the outer disc, 
creating a hot outflowing disc atmosphere with temperature 
$T\sim T_{\rm IC}=\int^{\infty}_0 h \nu L_{\nu} d\nu/4kL\sim 10^{7-8}$ K 
(Begelman et al. 1983a,b; Woods et al. 1996) whose  
ionisation parameter is expected to be primarily linked to the Spectral 
Energy Distribution (characterised by \tic) of the illuminating source. 

\begin{figure} 
\includegraphics[width=0.4\textwidth,height=0.495\textwidth,angle=-90]{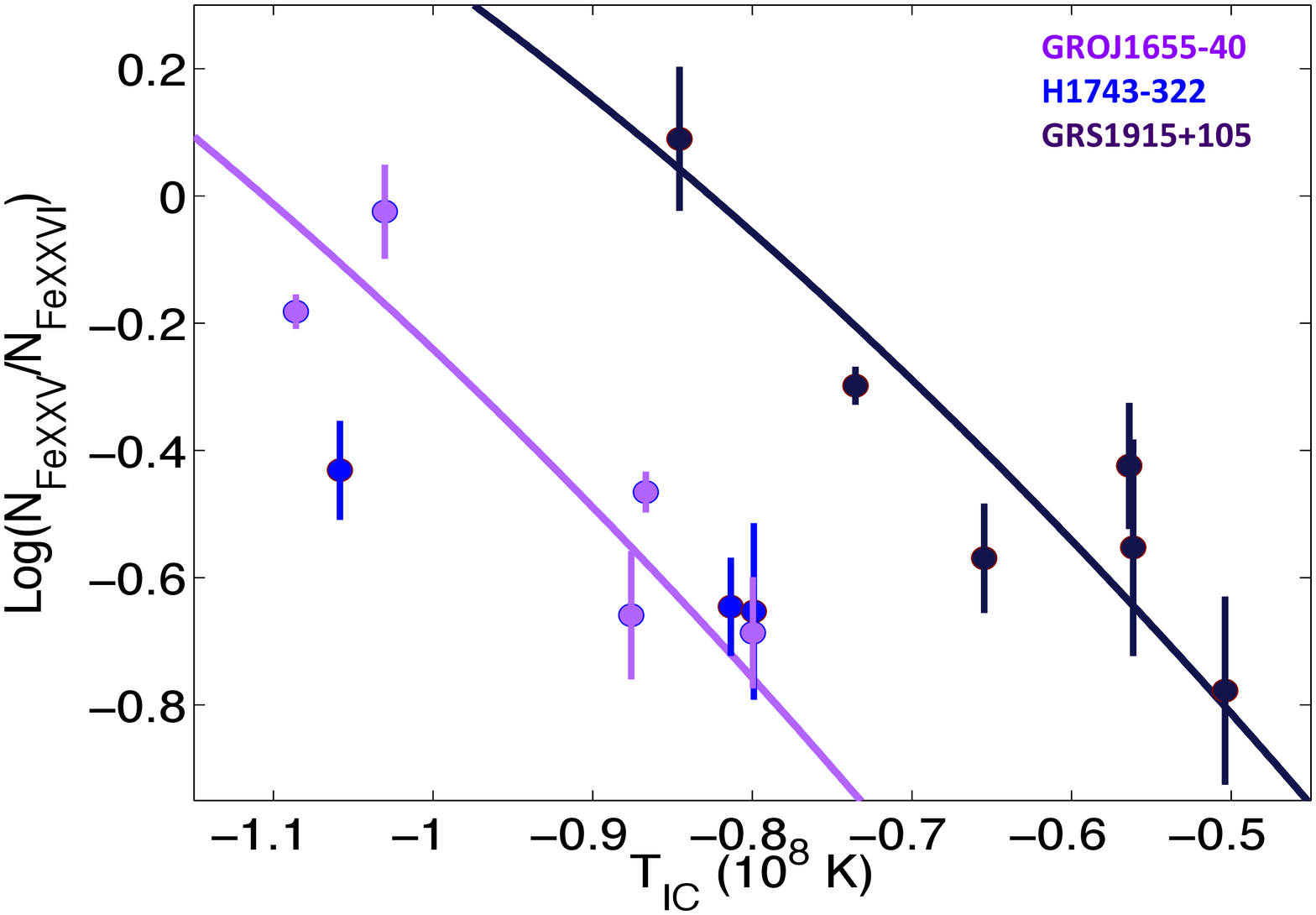}
\caption{\Fevc\ and \Fevs\ ion abundance ratio as a function of \tic. For each source the \Fevc/\Fevs\ ion abundance ratio is decreasing with \tic\ (which is a tracer of the spectral hardness). This suggests that the wind ionisation increases with spectral hardness, thus suggesting that ionisation effects play an important role here. The solid lines show the expected relation between ion ratios and \tic\ assuming linear a relation between $\xi$ and \tic\ ($\xi=\Xi\times T_{\rm IC}/1.92\times10^4$, where $\Xi=F_{\rm ion}/nkTc$, $F_{\rm ion}$ is the ionising flux between 1 and $10^3$ rydbergs; Krolik et al. 1981) and the ion fraction vs. $\xi$ as computed by Kallman \& Bautista (2001; see their Fig. 8) for an optically thin low density photo-ionised gas ($\Gamma=2$). 
}
\label{FeAb}
\end{figure}

Assuming that the \Fevc\ and \Fevs\ absorption lines are unsaturated 
and on the linear part of the curve of growth, we can estimate the \Fevc\ 
and \Fevs\ ion abundance (see formula 1 of Lee et al. 2002). 
Fig. \ref{FeAb} shows the \Fevc\ and \Fevs\ ion abundance ratio as 
a function of \tic\ for all the observations in which the two lines are detected. 
For all sources we systematically observe the lower ionisation states (\Fevc) 
at low \tic\ with the ratio \Fevc/\Fevs\ decreasing for harder spectral 
shapes (higher \tic), as expected if the ionisation parameter ($\xi$) increases 
linearly with \tic\ (see solid lines in Fig. 4).
This result suggests that, during the soft states, ionisation effects might play 
an important role in determining the properties of the wind. 
However, this conclusion is based on the ionisation balance for a low-density 
gas illuminated by a $\Gamma=2$ power law; in order to verify the 
disappearance of the wind in hard states as due to over-ionisation, 
a detailed study of the actual ionising spectra and wind densities would 
be required. Although important, this is beyond the scope of this paper.
\begin{figure} 
\includegraphics[width=0.38\textwidth,height=0.495\textwidth,angle=-90]{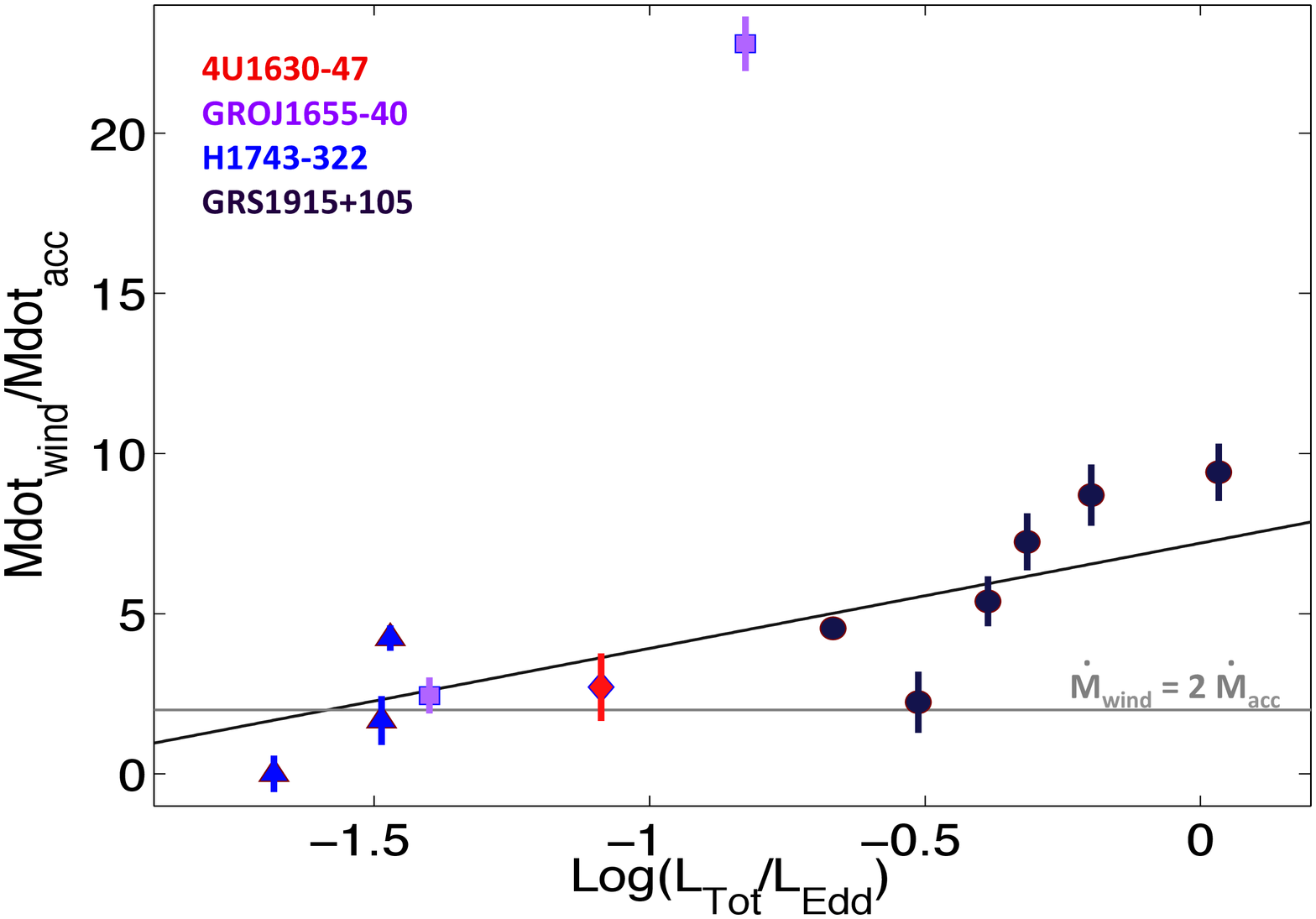}
\caption{$\dot{M}_{\rm wind}$ over $\dot{M}_{\rm acc}$ vs. luminosity. Apart from the one observation at the lowest luminosity, all detected winds carry away at least twice more mass than the one accreted into the central object. This implies that these winds are major players in the accretion phenomenon. The largest $\dot{M}_{\rm wind}$ is measured for the \gros\ observation during which a magnetically driven wind was detected (Miller et al. 2006a; Miller et al. 2008; Kallman et al. 2009). }
\label{FeKratio}
\end{figure}
Alternatively, the wind disappearance in the hard state might arise 
from the fact that illumination of the outer disc is critical for the production 
of Compton-heated winds or from some other phenomena 
(e.g. organisation of magnetic field) which is related to the accretion states. 
Irradiation only occurs when the outer disc subtends a larger solid 
angle than the inner flow. Thus, the formation of thermal winds 
might be prevented if harder states are associated with thick discs 
that, even if optically thin at the centre, have an optically thick 
region with H/R$\sim1$ or have a significant optical depth as 
seen from the outer disc (see also Neilsen et al. 2011b). Alternatively, 
if the disc ionisation instability 
is at work in these transient sources, the Compton radius of the 
wind might lie in a low-temperature, and thus un-flared, part of 
the outer disc (Dubus et al. 2001).

\section{Discussion}

How important are these winds for the accretion phenomenon? 
We estimate the wind mass outflow rate using the equation:
\begin{equation}
\dot{M}_{\rm wind}=4\pi R^2 n m_p v_{\rm out} \frac{\Omega}{4\pi}=4\pi m_p v_{out} \frac{L_X}{\xi} \frac{\Omega}{4\pi};
\end{equation}
where m$_p$ is the proton mass, v$_{\rm out}$ the wind outflow velocity 
and $\Omega$ is the solid angle subtended by the wind. 
\chandra\ observations provide reliable measurements 
of the outflow velocities, the detection of the wind in each soft state 
spectrum suggest a high filling factor, moreover we measured a wind 
opening angle of $\sim30^{\circ}$, thus, once estimated the ionisation 
parameter, we can measure the mass outflow rate and compare it to 
the mass inflow rate (assuming an efficiency $\eta=0.1$).
We estimate the ionisation parameter $\xi$ from the \Fevc\ / \Fevs\ 
ion ratio and assume the ion vs. $\xi$ distribution computed by 
Kallman \& Bautista (2001) and obtain values between Log($\xi$)$\sim3.5-4.2$. 
However, we caution the reader that these values might change significantly 
once (instead of assuming the ion abundances of Kallman \& Bautista 2001) 
the ion abundances vs. $\xi$ are computed using the properly tailored 
ionisation balances from the self consistent SED\footnote{
For example, under various assumptions about the gas density, different  
authors studying the same dataset found ionisation parameters that  
vary by $\sim2$ orders of magnitude (Miller et al. 2006a; 2008; Netzer 2006; 
Kallman et al. 2009). However, our estimated ionisation parameters here 
are within the typical range of measured values, and we believe they are 
useful for our purposes here.}. 
Figure \ref{FeKratio} shows that the mass outflow 
rates carried away by these winds are generally several times, up to 
10-20 times, higher than the mass accretion rates as found by 
Neilsen et al. (2011a). 
This indicates that these winds are fundamental components 
in the balance between accretion and ejection, and that disregarding 
such winds would mean overlooking the majority of the mass involved 
in the accretion phenomenon. Such massive winds suggest a higher 
mass transfer rate from the companion than is generally assumed. 
This might imply, for example, more rapid evolution of the binary 
orbit than we expect. 

Such winds should have a major impact on the 
physics of the inner accretion disc. For example, it is expected that 
the onset of the wind would reduce the local accretion rate. 
After a viscous time, 
this would modulate the accretion rate in the inner disc and, thus, 
the source luminosity, ultimately producing oscillations (Shield et al. 1986; 
Melia et al. 1991). 
Interestingly, a \suzaku\ observation caught \grsdn\ 
in transition from the hard ($\chi$ state) to the soft state (Ueda 
et al. 2010, but see also Neilsen et al. 2011) and, in agreement 
with this interpretation, both the rise of the wind and oscillations 
($\theta$ state) are observed. 
We note that Compton heated winds are predicted to be powerful 
only at luminosities higher than few per cent of the Eddington
limit. This corresponds to the only range of luminosities 
in which the soft states are observed, suggesting a strong connection. 
We speculate that the wind 
might have an effect in keeping the thin disc stable and the source 
firmly in the soft state, basically preventing the transition back to the 
hard state until the wind is not powerful anymore. This would lead
to a more or less constant luminosity for soft to hard state transitions,
which is observed (Maccarone 2002).
It is critical to establish in the future whether these winds 
really are driven by Compton heating and what their influence is on the inner 
accretion flow, to better understand how they are 
linked to the accretion state and/or the formation/suppression of the jet.

\section*{Acknowledgments}

The authors wish to thank L. Calvillo, M. Giustini, E. Koerding, 
C. Knigge and G. Dubus for discussion.
This research has made use both of data obtained from the \chandra, 
\xmm, \suzaku\ and \rxte\ satellites. GP acknowledges support via an EU 
Marie Curie Intra-European Fellowship under contract no. 
FP7-PEOPLE-2009-IEF-254279. MCB acknowledges support from NASA 
ATP grant NNX09AG02G and NSF grant AST-0907872.
JN acknowledges additional support from the National Aeronautics and  
Space Administration through the Smithsonian Astrophysical Observatory  
contract SV3-73016 to MIT for support of the Chandra X-ray Center,  
which is operated by the Smithsonian Astrophysical Observatory for and  
on behalf of the National Aeronautics Space Administration under  
contract NAS8-03060.

\end{document}